\documentclass[aps,amsmath,amssymb,twocolumn]{revtex4-2}

\usepackage{graphicx}
\usepackage{dcolumn}
\usepackage{footmisc}

\usepackage{comment}
\usepackage{bm}
\usepackage{xcolor}
\usepackage{float}
\usepackage{hyperref}

\newcommand{\aap}{Astronomy \& Astrophysics}

\newcommand{\apjl}{The Astrophysical Journal Letters}

\newcommand{\aj}{The Astronomical Journal}
\newcommand{\araa}{Annual Review of Astronomy and Astrophysics}

\begin{document}
\setlength{\abovedisplayskip}{5pt}
\setlength{\belowdisplayskip}{5pt}
\setlength{\abovedisplayshortskip}{5pt}
\setlength{\belowdisplayshortskip}{5pt}

\rightline{FERMILAB-PUB-25-0729-T}

\title{Dark Matter-Electron Interactions Alter the Luminosity and Spectral Index of M87}
\author{Abdelaziz Hussein}
\email{abdelh@mit.edu}
\affiliation{Department of Physics and Kavli Institute for Astrophysics and Space Research, Massachusetts Institute of Technology, 77 Massachusetts Ave, Cambridge MA 02139, USA}
\author{Gonzalo Herrera}
\email{gonzaloh@mit.edu}
\affiliation{Department of Physics and Kavli Institute for Astrophysics and Space Research, Massachusetts Institute of Technology, 77 Massachusetts Ave, Cambridge MA 02139, USA}
\affiliation{Harvard University, Department of Physics and Laboratory for Particle Physics and Cosmology, Cambridge, MA 02138, USA}
\affiliation{Center for Neutrino Physics, Department of Physics, Virginia Tech, Blacksburg, VA 24061, USA}

\begin{abstract}
We investigate the possibility that cosmic-ray electron cooling through dark matter–electron scatterings contributes to the low radiative efficiency observed in radio-loud galaxies such as M87. Light dark matter can scatter efficiently off electrons in M87, lowering the observed bolometric luminosity compared to astrophysical expectations. This consideration allows us to probe previously unexplored regions of the parameter space of dark matter-electron interactions. We further model the cosmic-ray electron distribution by numerically solving a diffusion equation along the jet and find that efficient dark matter–electron interactions can induce a flattening of the spectral index at different distances from the central supermassive black hole, in better alignment with radio observations from M87.
\end{abstract}

\maketitle

\textit{Introduction}—  The problem of quiescent black holes (BHs), where the observed bolometric luminosities are many orders of magnitude lower than the Eddington luminosity, remains an open problem in galactic astrophysics \cite{10.1093/mnras/277.1.L55}. It has been long assumed that the bolometric luminosity is solely a function of accretion rate along with some efficiency factors \cite{Kording_2006,Thorne_2025}. Thus, various accretion models have been developed to attempt to reconcile theory with observations. One such class is the Advection Dominated Accretion Flows (ADAF) \cite{1994ApJ...428L..13N},where
the bulk of the thermal energy is carried into the BH by the accreting gas as entropy rather than being radiated \cite{1994ApJ...428L..13N}. These models \cite{Reynolds_1996,1994ApJ...428L..13N,Ro_2023}, however, mischaracterize the observed flux and the spectral index distribution \cite{1998tx19.confE.400D}. Though the field has progressed to the use of sophisticated general relativistic magnetohydrodynamic (GRMHD) simulations, they often track only the dynamically important ion fluid, with no direct knowledge of the electrons available \cite{Davelaar_2019}. Furthermore, GRMHD simulations usually assume a thermal Maxwell-Juttner distribution \cite{EventHorizonTelescope:2019ggy}, which does not account for the non-thermal effects such as magnetic reconnection and turbulent dissipation, which will accelerate electrons
to the observed non-thermal power law distribution \cite{2024A&A...687A..88Z,Hoshino:2013pza}. These simple assumptions made regarding electron thermodynamics often limit the extent to which GRMHD simulations can be applied to observations of low-luminosity accreting BHs \cite{ressler2015electronthermodynamicsgrmhdsimulations,Chael_2018}. Thus, various phenomenological electron injection models have been developed to fit data \cite{Ro_2023}. Further extensions of these models have resulted in model degeneracies, where many different scenarios are able to fit the data \cite{2016ApJ...830...78L,Nagar_2001}.

\begin{figure*}[t!]
    \centering
    \includegraphics[width=0.4875\linewidth]{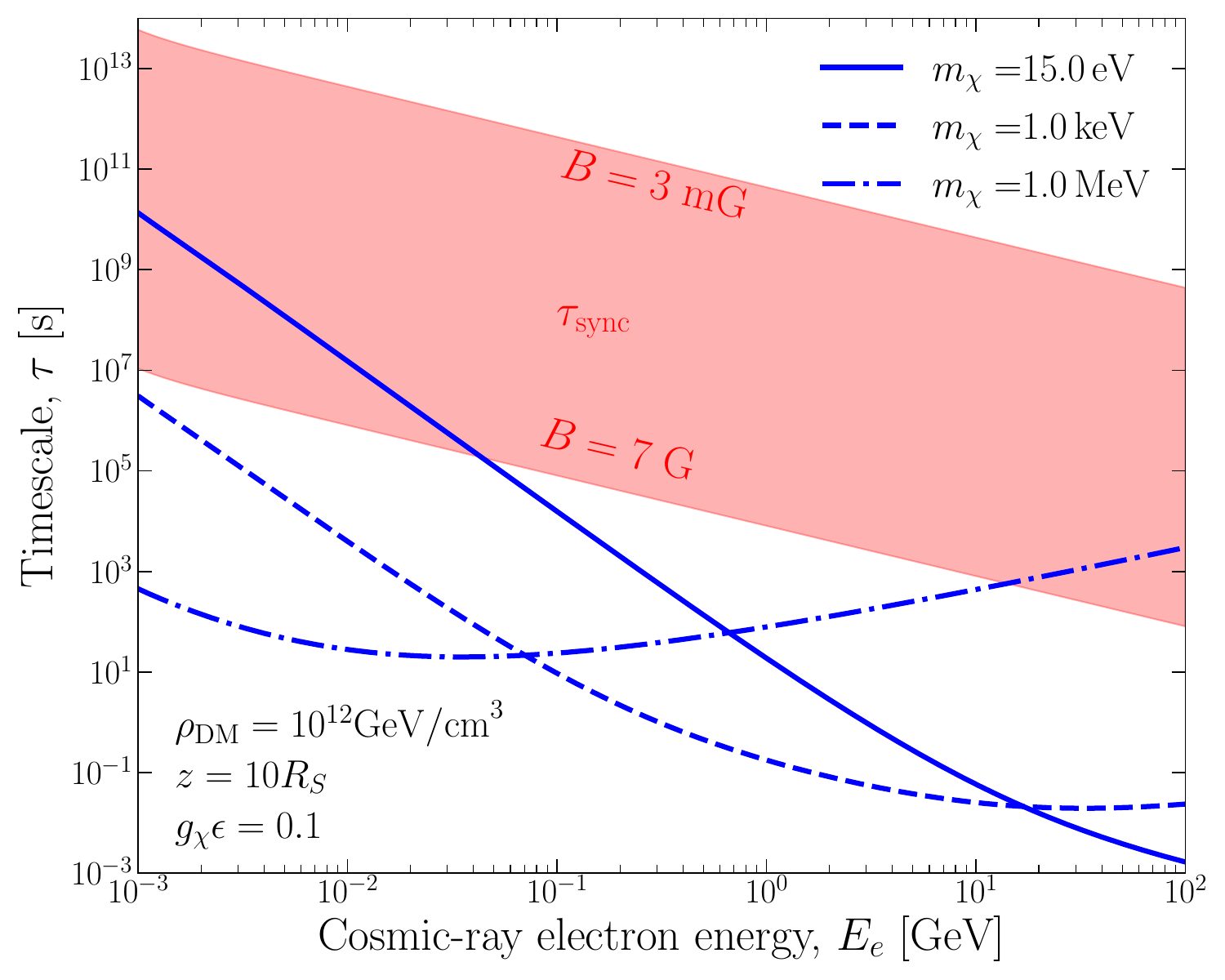}
    \includegraphics[width=0.485\linewidth]{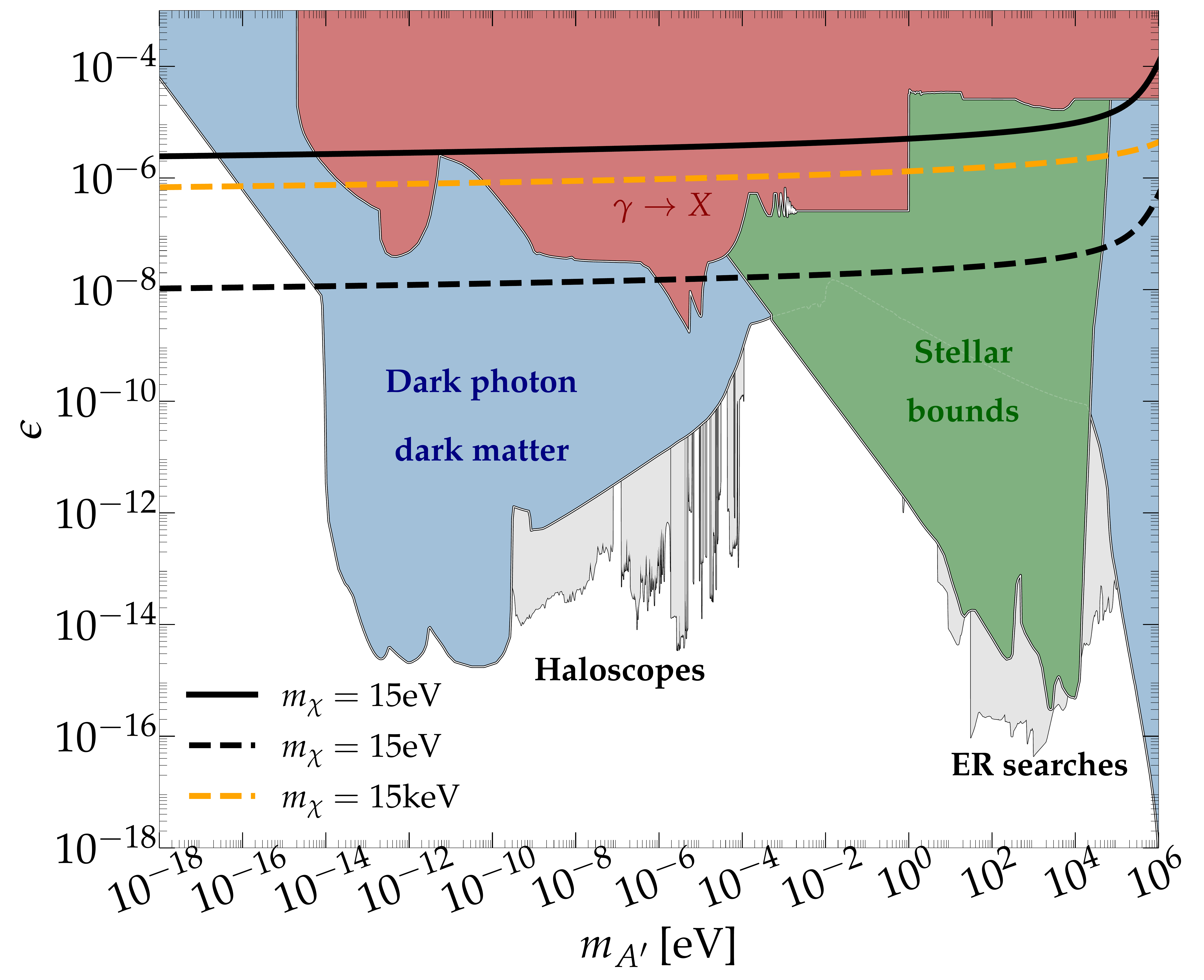}
    \caption{\textit{Left panel:} Cosmic ray electron timescales in M87. The red region corresponds to the synchrotron cooling timescale ($\tau_{\rm sync}$) given $B_{\text{jet}}$ in Eq. \ref{eq:B_jet} and the DM-electron scattering cooling timescales ($\tau_{\mathrm{\chi}e}$) are shown in blue, for different values of the DM mass, with $m_{A^{\prime}}=5$ MeV. \textit{Right panel:} Variety of constraints on the dark photon kinetic mixing $\epsilon$ versus dark photon mass from \cite{Caputo_2021}. The red region encloses all constraints that are based on photon to dark photon transitions with laboratory based searches (e.g. light-shining through wall experiments). In green, we show bounds arising from stellar cooling arguments, also relying on photon to dark photon transitions. In blue we show the bounds that rely on dark photons comprising the DM of the Universe. In gray, we show the direct detection bounds from devoted dark photon experiments, namely, SHUKET, WISPDMX, and Dark E-field Radio. The horizontal lines correspond to the constraints obtained in this work from M87 cooling, for various DM masses (in different colors), and fixed DM coupling $g_\chi = 0.1$. The dashed lines are derived using the magnetic field value $B_{\rm MAGIC}$, while the solid line uses $\langle B_{\rm VLBI}\rangle$ (see Appendix \ref{sec:appendix_B}). These bounds are evaluated at cosmic ray electron energy of $E_{e} = 10 \ \rm{GeV}$, and for an DM density $\rho_{\mathrm{DM}}(10R_s) = 10^{12} \rm{GeV/cm^3}$. }
    \label{fig:t_sync}
\end{figure*}

Another possibility is that the problem may not lie with the accretion model, but rather within overlooking the potential effect of particle dark matter (DM) on BH observations. Adiabatically-growing supermassive BHs (SMBH) can enhance the density of DM in their innermost vicinity, forming the so-called DM spike \cite{Quinlan:1994ed,Gondolo_1999}. Previous studies \cite{Herrera:2023nww,Gustafson:2024aom,Mishra:2025juk,Gorchtein:2010xa, Cermeno:2022rni, Kantzas:2025huu} have shown that in some models where the DM couples to electrons, DM is efficient in cooling energetic cosmic ray electrons around SMBHs. These works focused on deriving bounds on DM-electron interactions from either cooling arguments or modifications of the expected gamma-ray fluxes in certain sources.

In this \textit{Letter} we first show that DM-induced cosmic ray electron cooling can account for the lower bolometric luminosity of the extremely well-studied source M87. We assert that this cooling mechanism cannot lower the luminosity beyond the observed bolometric luminosity, allowing us to set competitive bounds on the parameter space of DM-electron interactions. We further track the impact of DM-electron interactions on the radio emission of M87 by solving a diffusion equation along the jet, finding the corresponding spectral index as a function of the distance from the BH. The additional cooling induced by DM generically flattens the spectral index, providing a novel method to constrain the particle nature of DM.

As a concrete example, we consider fermionic DM coupling to electrons via a dark photon mediator and set constraints on the sub-GeV mass region. Employing astrophysical observations to probe sub-GeV DM parameter space is particularly timely given recent direct detection results. The DAMIC-M \cite{DAMIC-M:2025luv} and PandaX-4T \cite{Zhang:2025ajc} collaborations recently achieved a milestone by extensively testing regions of parameter space consistent with the observed relic abundance via freeze-out with heavy mediators and freeze-in with ultralight or massless mediators \cite{DAMIC-M:2025luv, Krnjaic:2025noj, Cheek:2025nul, Zhang:2025ajc}. However, some scenarios remain poorly constrained by direct detection, notably Majorana DM, where the electron scattering cross section is velocity-suppressed ($v_{\chi} \sim 0.03 c$) in laboratory experiments. Since electrons are relativistic in radio-loud galaxy jets, our approach circumvents this velocity suppression, yielding bounds more stringent than those from direct detection.

\begin{figure*}
    \centering
    \includegraphics[width=0.49\linewidth]{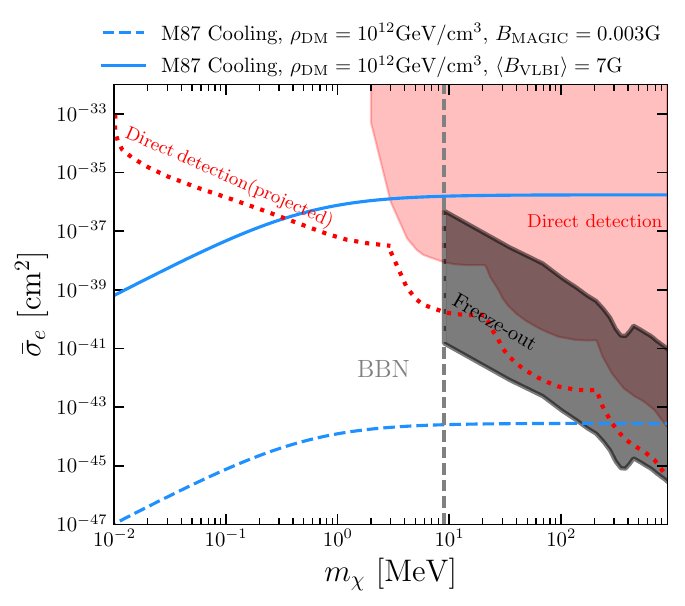}
    \includegraphics[width=0.49\linewidth]{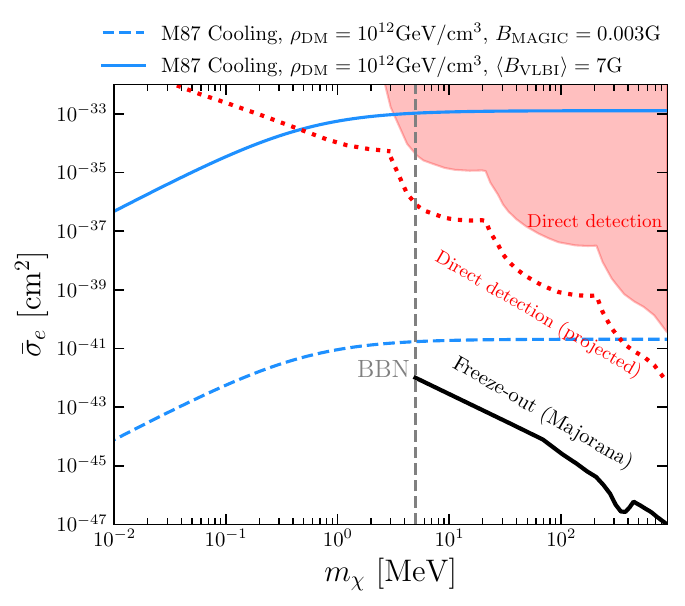}
    \caption{\textit{Left panel}: Upper limits on the non-relativistic Dirac DM-electron scattering cross section in terms of the DM mass. The blue band shows bounds from cosmic-ray electron cooling at M87 derived in this work, where the band accounts for two possible values of the magnetic field in the vicinity of the M87 BH inferred in the literature ($B=0.003-7$ G, see the Appendix \ref{sec:appendix_B} for more details). The DM density is taken at $10 R_S$, with value $\rho_{\rm DM}=10^{12}$GeV cm$^{-3}$; different values would rescale the bounds linearly (see the Appendix \ref{sec:appendix_DM} for a discussion on the DM density profile uncertainties). For comparison, we show as a vertical gray bound the lower limit on the DM mass obtained from measurements of the number of relativistic effective species ($N_{\rm eff}$) at Big Bang Nucleosynthesis \cite{Depta:2019lbe, Giovanetti:2021izc, Krnjaic:2019dzc}. The red region corresponds to bounds from DM-electron and DM-nucleus scatterings at current direct detection experiments. For masses below $m_{\chi} \lesssim 20$ MeV, the best bound comes from DAMIC-M \cite{DAMIC-M:2025luv}. For $ 20 \lesssim m_{\chi} \lesssim 200 $ MeV, the leading bound comes from electron recoils in PANDAX4T \cite{Cheek:2025nul, Zhang:2025ajc}. For masses above $m_{\chi} \gtrsim 200$ MeV, the leading bounds arise from nuclear recoils at CRESST-III \cite{CRESST:2019jnq}. The dotted red line is obtained from the combination of projected sensitivities in various future DM direct detection experiments. In particular, we take sensitivity projections from QROCODILE \cite{QROCODILE}, OSCURA \cite{aguilararevalo2022oscuraexperiment}, and XLZD \cite{XLZD:2024nsu}. The solid black band corresponds to the thermal relic abundance expectation in various models, see \textit{e.g} \cite{Krnjaic:2025noj, Hochberg:2014dra, Kuflik:2015isi, Graesser:2011wi, Lin:2011gj, Slatyer:2015jla}. \textit{Right panel:} Same plot as in the left panel, but for Majorana DM instead of Dirac DM. The direct detection bounds are relaxed compared to the Dirac case due to the velocity suppression of the scattering cross section $\sigma_{\rm \chi e} \propto v^2$. }
    \label{fig:sigma_bound}
\end{figure*}

\textit{Dark Matter Model and Distribution}— We consider an extension of the Standard Model (SM) by an additional $U(1)^{\prime}$ gauge symmetry, whose associated gauge boson $A^{\prime}$ kinetically mixes with the SM hypercharge boson with mixing strength $\epsilon$. The effective Lagrangian leading to interactions between DM and electrons reads:
\begin{align}
\mathcal{L} = & -\frac{1}{4}F_{\mu \nu}F^{\mu \nu} - \frac{1}{4}F^{\prime}_{\mu \nu}{F^{\prime}}^{\mu \nu} - \frac{\epsilon}{2}F_{\mu \nu}{F^{\prime}}^{\mu \nu} \notag \\
& + \frac{m_{A^{\prime}}^2}{2}A^{\prime}_{\mu}{A^{\prime}}^{\mu} + {A^{\prime}}^{\mu} g_{\chi} \bar{\chi} \gamma_\mu \chi
\label{eq:lagrangian}
\end{align}
where ${F^{\prime}}^{\mu \nu}=\partial^\mu {A^{\prime}}^{\nu}-\partial^\nu {A^{\prime}}^{\mu}$, and $\chi$ is the fermionic DM field. $g_{\chi}$ denotes the coupling strength of the DM to the dark photon $A^{\prime}$. The size of the kinetic mixing $\epsilon$ is generically model-dependent. If generated at the loop-level ($\epsilon \simeq g_{\chi}e/16\pi^2$), the combination of bounds from self-interactions and the thermal relic abundance severely restricts the allowed parameter space \cite{Cline:2024wja}. Here we do not impose any fixed relation among $g_{\chi}$ and $\epsilon$. 

From the described Lagrangian, we find that the differential DM-electron scattering cross section, in terms of the final DM kinetic energy $T_{\chi}$ and cosmic ray electron kinetic energy $T_e$, to be: 

\begin{align}
\frac{d \sigma}{d T_\chi} &= \frac{\epsilon^2 e^2g_\chi^2}{4 \pi\left(E_e^2-m_e^2\right)\left(m_{A^{\prime}}^2+2 m_\chi T_\chi\right)^2}\\ \notag
&\times \left[2 E_e^2 m_\chi-\left(m_e^2+m_\chi\left(2 E_e+m_\chi\right)\right) T_\chi+m_\chi T_\chi^2\right]
\end{align}

where $m_e$ is the electron mass, $m_\chi$ denotes the DM mass, and $m_{A^{\prime}}$ denotes the dark photon mass.

Notably, DM-electron interactions cool the cosmic ray electrons as the electrons are relativistic in the jet, where the Lorentz factor is $\gamma_e \sim 1000$, while the DM in the region of interest is comparatively at rest, $\gamma_{\chi} \sim 0.5$ \cite{sabarish2025accretionselfinteractingdarkmatter}. From the differential cross section, the DM-electron cooling timescale can be found as:
\begin{equation}
\tau_{\mathrm{\chi}e}=\left[-\frac{1}{E_e}\left(\frac{d E_e}{d t}\right)_{\mathrm{\chi}e}\right]^{-1},
\end{equation}
with
\begin{equation}
\left(\frac{d E_e}{d t}\right)_{\chi e}=-\frac{ \rho_{\rm DM} }{m_{\chi}} \int_0^{T_{\chi}^{\max }} d T_{\chi} T_{\chi} \frac{d \sigma_{\chi e}}{d T_{\chi}}.
\label{DM-e power}
\end{equation}
The maximum DM kinetic energy allowed by the collision kinematics is dictated by:
\begin{equation}
    T^{\rm{max}}_{\chi} = \frac{T_e^2+2m_eT_e}{T_e + \frac{(m_e+m_{\chi})^2}{2m_{\chi}}}. 
\end{equation}

Another crucial aspect of our set-up is the DM density profile of M87. We consider a DM spike arising from adiabatic growth of the central SMBH \cite{Gondolo_1999})
\begin{align}\label{eq:spike}
	\rho_{\rm DM}(r) = \rho_{R} \, g_{\gamma}(r)\, \Big(\frac{R_{\rm sp}}{r}\Big)^{\gamma_{\rm sp}}\;,
\end{align}
and where $R_{\rm sp}=\alpha_{\gamma}r_0(M_{\rm BH}/(\rho_{0}r_{0}^{3})^{\frac{1}{3-\gamma}}$ is the size of the spike,  and $\gamma_{\rm sp}=\frac{9-2\gamma}{4-\gamma}$ is the steepness of the spike. Further, the factor $g_{\gamma}(r) \simeq (1-\frac{2R_{S}}{r})$ \cite{Sadeghian:2013laa}, with $R_S$ denoting the Schwarzschild radius, while $\rho_{\mathrm{R}}=\rho_0\left(\frac{R_{\mathrm{sp}}}{r_0}\right)^{-\gamma}$. For M87, we find $R_{\rm{sp}} \approx 4\times10^5R_s$ for $M_{\rm{BH}} = 6.5 \times10^9M_\odot,\alpha_\gamma = 0.1, \ r_0 = 20 \rm{\  kpc}, \gamma = 1, \rho_0 = 2.5 \, \rm{GeV}/\rm{cm}^3$ \cite{PhysRevD.92.043510, EventHorizonTelescope:2019ggy}. This prescription for the DM density profile may be subject to modifications due to a variety of astrophysical processes, which we discuss in Appendix \ref{sec:appendix_DM}.

 \begin{figure*}[t!]
  \centering
  \includegraphics[width=0.49\linewidth]{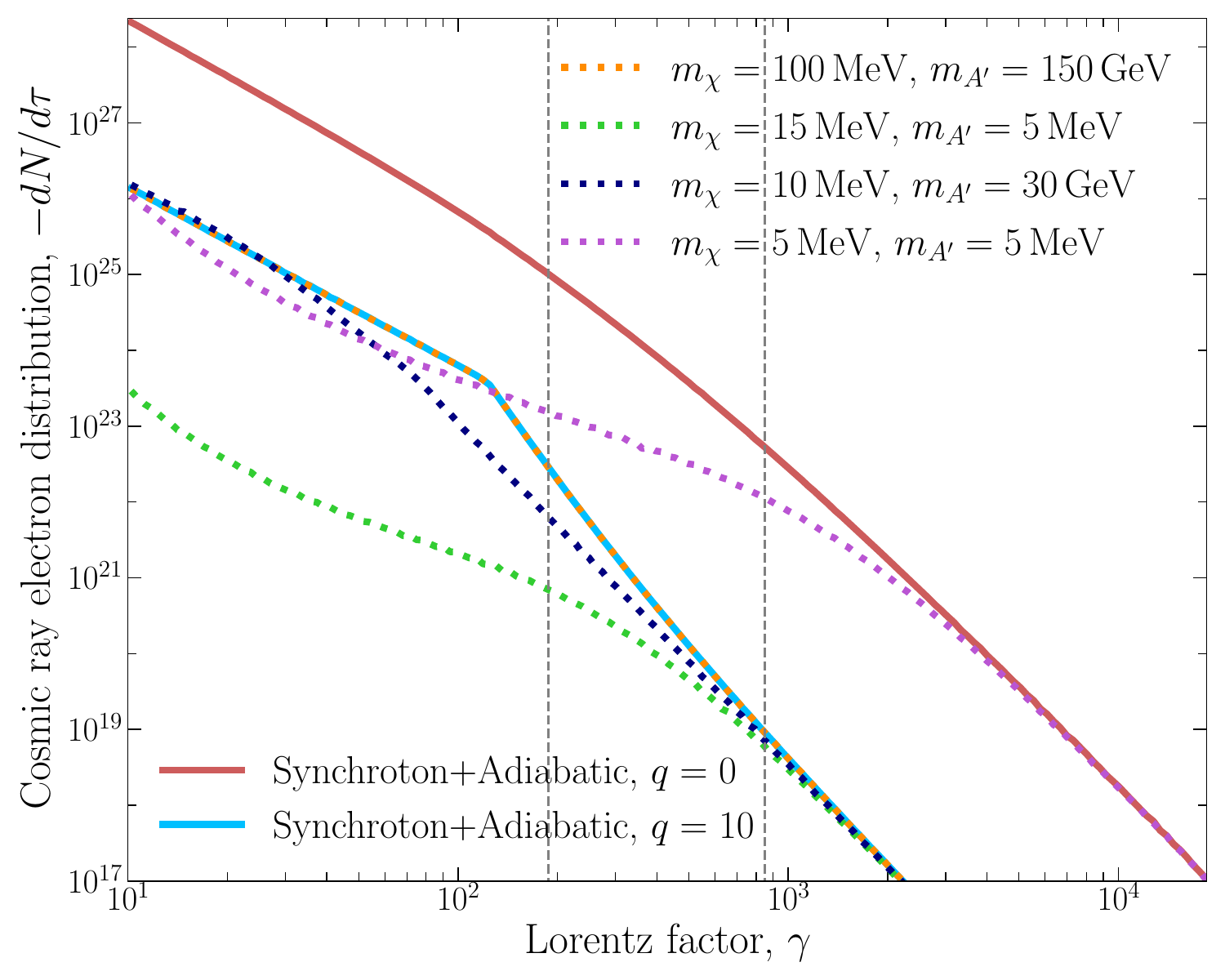}
\includegraphics[width=0.49\linewidth]{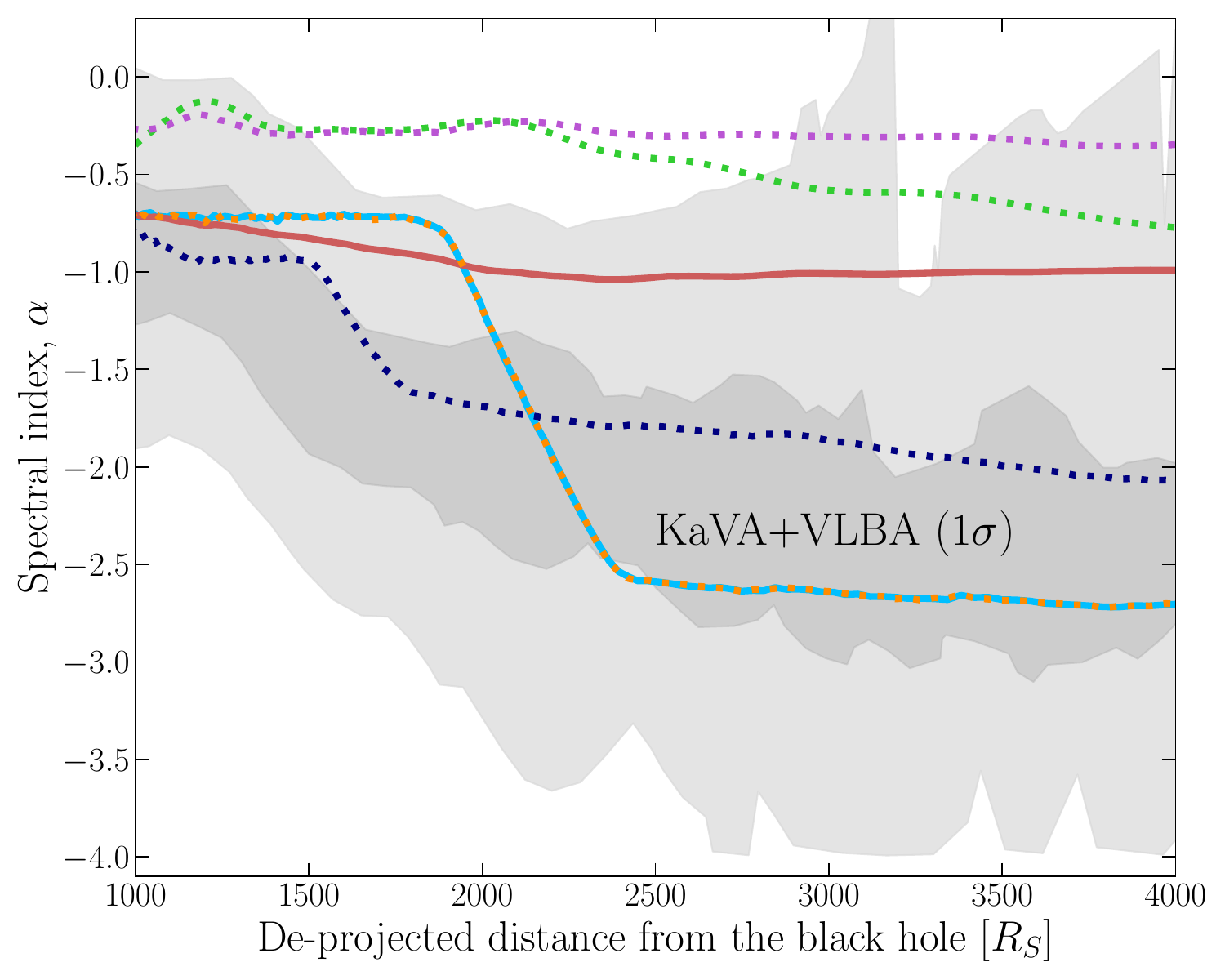}
    \caption{\textit{Left panel:} Cosmic ray electron distribution as a function of energy, obtained by solving Eq. \ref{eq:transfer_eq}. The solid lines represent the cosmic ray-electron distribution obtained from adiabatic and synchroton cooling, in the absence of DM-electron interactions. The dotted lines corresponds to the evolution of the cosmic ray electron distribution when including DM-electron interactions. We set the product of couplings $g_\chi \epsilon =0.1$. The lines $q =0$ and $q =10$ are in agreement with Ref. \cite{Ro_2023}. Depending on the configuration, DM can either leave the distribution unaffected or flatten it. This flattening occurs in the regions where DM cooling is more efficient than SM cooling. The vertical dashed lines represent the turn over energies where SM cooling becomes efficient in depleting the cosmic ray electron flux. \textit{Right panel:} Spectral index $\alpha$ versus de-projected distance from the SMBH in units of $R_S$, with the legend as in the left panel. We show in solid lines the simulated spectral index obtained in absence of DM-electron interactions for $q=0$ and $q=10$, in agreement with \cite{Ro_2023}. The dotted lines correspond to the modified spectral index in presence of DM-electron scatterings for different choices of parameters. For efficient DM-electron induced cooling, the spectral index flattens more than expected from conventional astrophysical mechanisms. For comparison, the shaded gray region shows the range of spectral index observations inferred from a combination of KaVA and VLBA data \cite{Niinuma:2014lga, 2011Natur.477..185H,Ro_2023}, together with the $1\sigma$ error band.}
    \label{fig:spectral_index}
\end{figure*}

\textit{Cosmic Ray Electron Cooling Timescales}— The cooling timescale of electrons induced by synchrotron radiation is given by \cite{1986rpa..book.....R}:

\begin{equation}
\tau_{\mathrm{syn}}=\Big[\frac{4 c \sigma_T}{3 E_{e}} \frac{B^2}{8 \pi} \gamma_{\mathrm{e}} \beta_{\mathrm{e}}^2\Big]^{-1},
\end{equation}

where $\sigma_{\rm T}\simeq 6.65 \times 10^{-25}$cm$^2$ is the Thomson cross section, $B$ denotes the magnetic field, $\gamma_e$ is the Lorentz factor of the electron, and $\beta_e$ is the velocity of the electron. To assess the impact of DM-electron interactions on the bolometric luminosity of M87, we consider two representative values of the magnetic field inferred from independent methods (see Appendix \ref{sec:appendix_B} for details). As a conservative value, we take an average of various measurements obtained with Very Long Baseline Interferometry (VLBI) at a distance $z = 10R_s$ from the SMBH, with value $\langle B_{\rm VLBI} \rangle=7 \rm{G}$ ~\cite{Kino:2014dua, Kino:2015dha, Hada_2012, Hada_2016, Acciari_2010, Kim:2018hul, 2014Natur.510..126Z, EventHorizonTelescope:2019dse, Jiang_2021}. As an aggressive estimate, we take the inferred value from the Spectral Energy Distribution (SED) fits at the MAGIC telescope ~\cite{MAGIC:2020gbb}, of $B_{\rm MAGIC}=0.003$G.

\textit{Constraints}—
We consider the scenario in which DM–electron scattering accounts for the low Eddington ratio observed in M87, allowing us to relate the luminosity ratios to the corresponding cooling timescales as follows (see Appendix \ref{sec:optical_depth_argument}):
\begin{equation}
     \log \left(\frac{L_{\rm bol}}{L_{\rm edd}} \right) = -\frac{\tau_{\rm sync}}{\tau_{\mathrm{\chi}e}}.
\end{equation}

The ratio of the Eddington luminosity, $L_{\rm edd} \approx 8.45 \times 10^{47}\ {\rm erg\ s}^{-1}$, to the bolometric luminosity of the core, $L_{\rm bol} \sim 10^{42}\ {\rm erg\ s}^{-1}$  \cite{Reynolds_1996,prieto_2016}, yields:

\begin{equation}
     \log\!\left(\frac{L_{\rm bol}}{L_{\rm edd}}\right) = -\frac{\tau_{\rm sync}}{\tau_{\mathrm{\chi}e}} \simeq -9
\end{equation}
Consequently, we consider that DM-e scattering cross sections leading to ${\tau_{\rm{sync}}}/{\tau_{\mathrm{\chi}e}} \ge C$ are ruled out, where we conservatively set $C = 10$.

We plot $\tau_{\rm{sync}}$ and $\tau_{\mathrm{\chi}e}$ for a fixed product of couplings $g_\chi \epsilon = 0.1$, mediator mass $m_{A'} = 5 \ \rm{MeV}$, $\rho_{\rm DM}(10R_s) = 10^{12}  \ \rm{GeV/cm}^3 $ in Fig. \ref{fig:t_sync}. We notice that as $m_{\chi}$ decreases, $\tau_{\mathrm{\chi}e}$ becomes lower (i.e. DM cooling is more efficient). Similarly, for lower values of $m_{A'}$ and larger $g_\chi \epsilon$, DM cooling is more efficient. This is mainly due to the total number DM scatterings increasing with lower  $m_{\chi}$.

 We show constraints on $\epsilon$ versus $m_{A^{\prime}}$ in the right panel of Fig. \ref{fig:t_sync}, where we have taken  $g_\chi \epsilon = 0.1$, $E_{e} = 10 \ \rm{GeV}$, and $\rho_{\rm DM}(10R_s) = 10^{12} \ \rm{GeV/cm^3}$. We focus on very light DM states, which would require non-thermal production mechanisms in the early Universe to avoid constraints from  big-bang nucleosynthesis (BBN) and the CMB.  We plot the bound for $m_\chi = 15 \ \rm eV$ and $15 \ \rm keV$ to stay within the Tremaine-Gunn bound \cite{1979PhRvL..42..407T}. The dashed lines are calculated using $B_{\rm MAGIC}$, while the solid line is calculated using $\langle B_{\rm VLBI}\rangle$. We further plot current leading constraints from dark photon experiments, taken from Ref. \cite{Caputo_2021}. We note that the bound behaves as a power-law in dark photon mass that plateaus when $m_{A^\prime} \simeq \sqrt{\frac{2}{5}m_\chi T^{\rm{max}}_\chi}$. Our bounds are strongest for light mediator masses where we constrain open parameter space. The value of $\epsilon$ at which the plateus curve decreases as $m_\chi$ decreases, and thus we obtain our strongest constraints for lightest DM masses.

We further explore the effects of MeV scale DM on cosmic ray electrons, where thermal freeze-out remains untested in certain regions of parameter space, and some direct detection experiments on Earth are able to place meaningful constraints. In particular, we consider both the cases of Dirac and Majorana fields for $\chi$ in Eq. \ref{eq:lagrangian}, and the limit $m_{A^{\prime}} \gg \alpha m_e$. In this case, the non-relativistic DM-electron scattering cross section for the Dirac case reads \cite{Essig:2011nj}:
\begin{equation}
\bar{\sigma}_{e} \simeq\frac{4 \mu_{\chi e}^2 \alpha g_{\chi}^2 \varepsilon^2}{\pi m_{A^{\prime}}^4}.
\end{equation}
and for the Majorana case is \cite{Berlin:2018bsc}:
\begin{equation}
\bar{\sigma}_{e} \simeq\frac{6 \mu_{\chi e}^2 \alpha g_{\chi}^2 \varepsilon^2}{\pi m_{A^{\prime}}^4}v^2,
\end{equation}
where $v$ is the relative velocity of the DM-electron system. 

In Fig. \ref{fig:sigma_bound} we show bounds on $\bar{\sigma}_e$ as a function of $m_{\chi}$ (left panel for Dirac DM, right panel for Majorana DM) from cosmic-ray electron cooling in M87. These bounds are shown as two blue lines encompassing the uncertainty on the magnetic field along the jet of M87. The red shaded region shows current limits from direct detection experiments, while the dotted red line indicates projected bounds from a combination of future experiments. The lower bound obtained from measurements of the number of relativistic species $N_{\rm eff}$ at the time of Big Bang nucleosynthesis is shown using the dotted gray line. Finally, the black shaded region in the left plot indicates parameter space able to yield thermal relic DM, while the black line in the right plot illustrates the thermal relic target for Majorana DM. 
For both Dirac and Majorana cases, M87 cooling bounds using $\langle B_{\rm VLBI}\rangle$ exceed direct detection constraints at $m_{\chi} \le 6$ MeV. Using $B_{\rm MAGIC}$, the M87 constraints are approximately 2 (4) orders of magnitude stronger than upcoming direct detection experiments at $m_\chi = 10 \rm \ MeV$ for the Dirac (Majorana) case. These results highlight the potential constraining power of this method and the limitations due to the uncertainty in the magnetic field measurements.

\textit{Electron Distribution and Spectral Index Modeling}— To account for the effects of DM-electron scatterings on the cosmic ray electron distribution, we first model the cosmic ray electron distribution in the absence of DM-induced effects by reproducing the results from \cite{Ro_2023}. In the frame co-moving with
the jet’s flow, the transfer diffusion equation is written as:
\begin{equation}
    \frac{N(\gamma,\tau)}{\partial \tau}+(\nabla \cdot v)N(\gamma,\tau)+\frac{\partial}{\partial \gamma}[b(\gamma,\tau)N(\gamma,\tau)] = Q(\gamma,\tau).
\end{equation}
Here $N(\gamma,\tau)$ is the number density of the non-thermal electrons as a function of energy $\gamma$ and time $\tau$. $v$ is the velocity of the system, $Q$ is the nonthermal electron injection rate and $b$ is the energy loss rate. Standard model processes, namely adiabatic and synchrotron losses are traditionally considered the dominant energy loss processes, where the energy loss terms can be written as:
\begin{equation}
    b(\gamma,\tau)_{\rm{SM}} = \frac{d\gamma}{d\tau} = -b_{\rm adi}\gamma-b_{\rm sync}\gamma^2.
\end{equation}
The adiabatic energy loss term reads:
\begin{equation}
b_{\rm adi} = \frac{1}{R}\frac{dR}{d\tau} = \frac{0.56}{z}\Gamma(z) \beta(z),
\end{equation}
and the synchroton energy loss term reads
\begin{equation}
b_{\rm sync} = \frac{4}{3}\frac{\sigma_T}{m_ec}\frac{B^2}{8\pi},
\end{equation}
where $R$ is the jet radius in the comoving fame and $B$ is the magnetic field strength in the comoving frame. 

Accounting for DM cooling, the total cooling rate now reads: \begin{equation}
    b = b_{\rm{SM}}+b_{\rm{DM}}
\end{equation} 
where
\begin{equation}
b_{\rm DM} = \frac{1}{m_e} \left (\frac{dE}{dt}\right )_{\chi e}.
\end{equation}
$ (dE/dt)_{\chi e}$ is given in Eq. \ref{DM-e power}, and we track the distance-dependent DM density profile according to Eq. \ref{eq:spike}. Taking $\nabla \cdot v =(3/R)(dR/d\tau)=3b_{\rm adi}$, we arrive at the simplified transfer equation:
\begin{equation}\label{eq:transfer_eq}
    \frac{d N(\gamma,\tau)}{d \tau} = Q(\gamma,\tau)- 3b_{\rm adi}N(\gamma,\tau)-\frac{\partial b(\gamma,\tau)}{\partial \gamma}N(\gamma,\tau) .
\end{equation}
We take the non-thermal electron injection model and associated parameters from \cite{Ro_2023}
\begin{equation} \label{eq:Q}
    Q(\gamma,\tau) = Q_0 \gamma^p \left(\frac{z}{z_i}\right)^{-q},
\end{equation}
where $Q_0,p,q,z_i$ are variables fitted to the data.

In this calculation, we take the magnetic field along the jet according to \cite{Ro_2023}, where the magnetic field tracks the distance-dependent changes in the synchrotron spectrum of the jet from high-resolution very long baseline interferometry observations. These observations were made using the Korean Very Long Baseline Interferometry Network (KVN), VERA Array (KaVA), and the Very Long Baseline Array (VLBA) \cite{Niinuma:2014lga, 2011Natur.477..185H}. The strength of the magnetic field evolves as a function of distance to the central BH $z$ as
\begin{equation}
    B_{\text{jet}}(z) = (0.3-1\text{G})\Big{(}\frac{z}{900R_s}\Big{)}^{-0.72} ,\quad 900R_s<z<4500R_s,
    \label{eq:B_jet}
\end{equation}

We numerically solve Eq. \ref{eq:transfer_eq} and plot our results in Fig \ref{fig:spectral_index}. In the left panel of the Figure, we consider two benchmark scenarios: $q =0$ represents the canonical electron energy injection function, which does not fit the observed spectral index well, and $q = 10$ represents the proposed phenomenological model in \cite{Ro_2023}. We further plot three different values of the DM mass and mediator mass for each $q$-value. For instance, for $m_{\chi} = 100 \, \rm{MeV}$, $m_{A^{\prime}} = 150 \, \rm{MeV}$, $g_{\chi}\epsilon=0.1$ and $q=10$, the electron number density is not affected. However, for $m_{\chi} = 15 \, \rm{MeV}$, $m_{A^{\prime}} = 5 \, \rm{MeV}$, $q=10$, DM-electron interactions appear to have a strong impact on the cosmic-ray electron distribution, depleting the cosmic ray electron flux at low-energies. 

Using the electron energy distribution, we create spectral index maps as a function of deprojected distance from the central SMBH. For a given energy range, the electron number distribution can be modeled as $N(\gamma) \propto \gamma^{-p}$, where $\gamma$ is the Lorentz factor, and $p$ is the particle distribution index. The spectral index is then related to the particle distribution index as $\alpha=(p+1)/{2}$\cite{1986rpa..book.....R}. Changes in the flux at two different frequencies suffices to calculate $\alpha$ and compare against observations. In particular, we compare to observations from VLBA and KaVA reported in \cite{Ro_2023} using 22 GHz and 43 GHz images. These bands are chosen to avoid the opacity due to molecular absorption in the troposphere \cite{Radio_review}. Thus, we calculate $\alpha_{22-43\text{GHz}}$ from 
$p_{22\text{--}43\,\text{GHz}} = \dfrac{\log [N(\gamma_{22})/N(\gamma_{43})]}{\log [\gamma_{22}/\gamma_{43}]}$.

The results are shown in the right panel of Fig. \ref{fig:spectral_index}, where the gray band is the range of spectral index observations inferred from a combination of KaVA and VLBA data, together with the $1\sigma$ error band on those measurements \cite{Niinuma:2014lga, 2011Natur.477..185H,Ro_2023}. We show that for $m_{\chi} = 100 \, \rm{MeV}$, $m_{A^{\prime}} = 150 \, \rm{MeV}$, $q=10$, the spectral index does not change, while for the other configurations, we observe that DM flattens the spectral index. For $m_{\chi} = 15 \, \rm{MeV}$, $m_{A^{\prime}} = 5 \, \rm{MeV}$, $q=10$ and $m_{\chi} = 5 \, \rm{MeV}$, $m_{A^{\prime}} = 5 \, \rm{MeV}$, $q=10$, the predicted spectral index is ruled out by observations. While for  $m_{\chi} = 10 \, \rm{MeV}$, $m_{A^{\prime}} = 30 \, \rm{MeV}$, $q=10$, the predicted spectral index is a better fit to the data, falling mostly within the $1\sigma$ band.  These outcomes depend on how the DM cooling rate compares to the adiabatic and synchroton cooling rates for a given set of DM masses, mediator masses and couplings. 

We emphasize that the impact of DM-electron interactions on the spectral index presents a novel method to constrain DM-electron interactions. The constraint power of this method, however, relies on understanding the standard model processes that are shaping the electron distribution. Advances in understanding the standard model processes accelerating the electrons through particle-in-cell simulations \cite{Galishnikova_2023,chen2025introducingaperturegpubasedgeneral,hakobyan2025reconnectiondrivenflaresm87protonsynchrotron} and theoretical work will allow the future use of our method to provide leading constraints on DM-electron interactions.

\textit{Summary}— In this \textit{Letter}, we studied how DM-electron scatterings can account for the low bolometric luminosity observed for M87 relative to its Eddington luminosity. If we assert that this cooling mechanism cannot lower the luminosity beyond the observed bolometric luminosity, we are able to set competitive bounds on the parameter space of DM-electron interactions via a dark photon mediator. We further numerically solve the transfer equation in \ref{eq:transfer_eq} to track the effects of DM cooling on the electron energy distribution, and additionally calculate the spectral index at different locations along the jet. We find that efficient DM cooling can flatten the spectral index more than expected from adiabatic and synchroton cooling processes. Thus, we present for the first time spectral index observations at different distances from the central SMBH as a tool to constrain DM-electron interactions.

We foresee our work charting a new path to probe DM models via their impact on the cosmic ray electron distribution from radio-loud galaxies. Future work should extend this framework beyond M87 to ensembles of radio-loud AGN, where correlations between luminosity suppression, spectral index evolution, and host-galaxy properties could enhance the sensitivity of these constraints. Dedicated particle-independent methods to determine the DM density profile in these environments (\textit{e.g} \cite{Sharma:2025ynw}) will also be crucial in narrowing down the DM-cooling uncertainties. Furthermore, incorporating simultaneous electromagnetic observations at different wavelengths could allow us to better disentangle DM–induced cooling from conventional astrophysical processes, and to test the universality of the mechanism across diverse environments.

\bigskip
{\bf Acknowledgements.}
We would like to thank Hyunwook Ro, Motoki Kino, Uddipan Banik, Lina Necib, Nick Ekanger, Shunsaku Horiuchi and Elena Pinetti for useful discussions. The authors acknowledge the MIT Office of Research Computing and Data for providing high performance computing resources that have contributed to the research results reported within this paper. The work of GH is supported by the Neutrino Theory Network Fellowship with contract number 726844, and by the U.S. Department of Energy under award number DE-SC0020262. This manuscript has been authored by FermiForward Discovery Group, LLC under Contract No. 89243024CSC000002 with the U.S. Department of Energy, Office of Science, Office of High Energy Physics.
\bibliography{References}

\clearpage
\appendix
\onecolumngrid
\section{Uncertainties on the dark matter distribution in M87}\label{sec:appendix_DM}

The results obtained in the main text are based on a dark matter density profile corresponding to a spike with initial index $\gamma=1$, as described in Eq. \ref{eq:spike}. 
Although the formation of a dark matter spike is expected under adiabatic growth, its survival over sufficiently long timescales depends on additional assumptions.

For example, dark matter self-annihilations could deplete the spike of M87, with plateau density given by $\rho_{\rm sat}=m_{\rm DM}/(\langle \sigma v \rangle t_{\rm BH})$, where $t_{\rm BH} \simeq 10^{10}$yr. The depletion is negligible if the thermally averaged annihilation cross section lies roughly below the thermal relic freeze-out value expected in some models, $\langle \sigma v \rangle \lesssim 10^{-26}\mathrm{cm}^{3}\mathrm{s}^{-1} \lesssim 0.23 \times \langle \sigma v \rangle_{\rm freeze-out}$ \cite{Lacroix:2015lxa, Steigman:2012nb}, such as for asymmetric dark matter.

Additional astrophysical processes may further deplete the spike, with the two dominant effects likely being stellar heating by baryons \cite{Merritt:2006mt} and mergers \cite{Merritt:2002vj}. In the case of efficient stellar heating, the dark matter spike may relax via the Bahcall–Wolf mechanism to a shallower configuration with a spike index $\gamma_{s}=3/2$ \cite{Merritt:2006mt}. The effect is controlled by the relaxation timescale \cite{2008gady.book.....B}

\begin{equation}
t_r(r) \simeq \frac{0.34 \sigma^3}{G^2 m_{\star} \rho_{\star}(r) \ln \Lambda},
\end{equation}
where we can assume for most stars $\sigma \sim 200$ km s$^{-1}$, $m_{\star} \sim M_{\odot}$, and the Coulomb logarithm in galactic nuclei is $\mathrm{ln} \Lambda \sim 15$ \cite{gualandris2007dynamicssupermassiveblackholes}. Then we can estimate the stellar density required to relax the system on smaller timescales than it's formation timescale $t_r \lesssim t_{\rm BH}$, for $r_h \sim 900-4500 R_s$. We then find

\begin{equation}\label{eq:stellar_density}
\rho_{\star} \gtrsim 9 \times 10^5 \frac{M_{\odot}}{\mathrm{pc}^3}\left(\frac{\sigma}{200 \mathrm{~km} \mathrm{~s}^{-1}}\right)^3\left(\frac{15}{\ln \Lambda}\right)\left(\frac{M_{\odot}}{m_{\star}}\right)\left(\frac{10^{10} \mathrm{yr}}{t_{\rm BH}}\right).
\end{equation}

The stellar density in M87 is likely smaller than the value we obtained in Eq. \ref{eq:stellar_density}. Observations of the central starlight luminosity of M87 with the Hubble Space Telescope indicate $L_{\star} \gtrsim 10^{3}L_{\odot} \mathrm{pc}^{-3}$ for $r<10$pc \cite{1992AJ....103..703L}. Further assuming a mass to light ratio of $M/L \sim 5$, we find a stellar density of $\rho_{\star} \simeq 10^{4} M_{\odot}$pc$^{-3}$, a couple of orders of magnitude below the requirement of Eq. \ref{eq:stellar_density}. We thus conclude that stellar heating is unlikely to modify the dark matter spike fo M87.

\begin{figure}[H]
    \centering
    \includegraphics[width=0.7\linewidth]{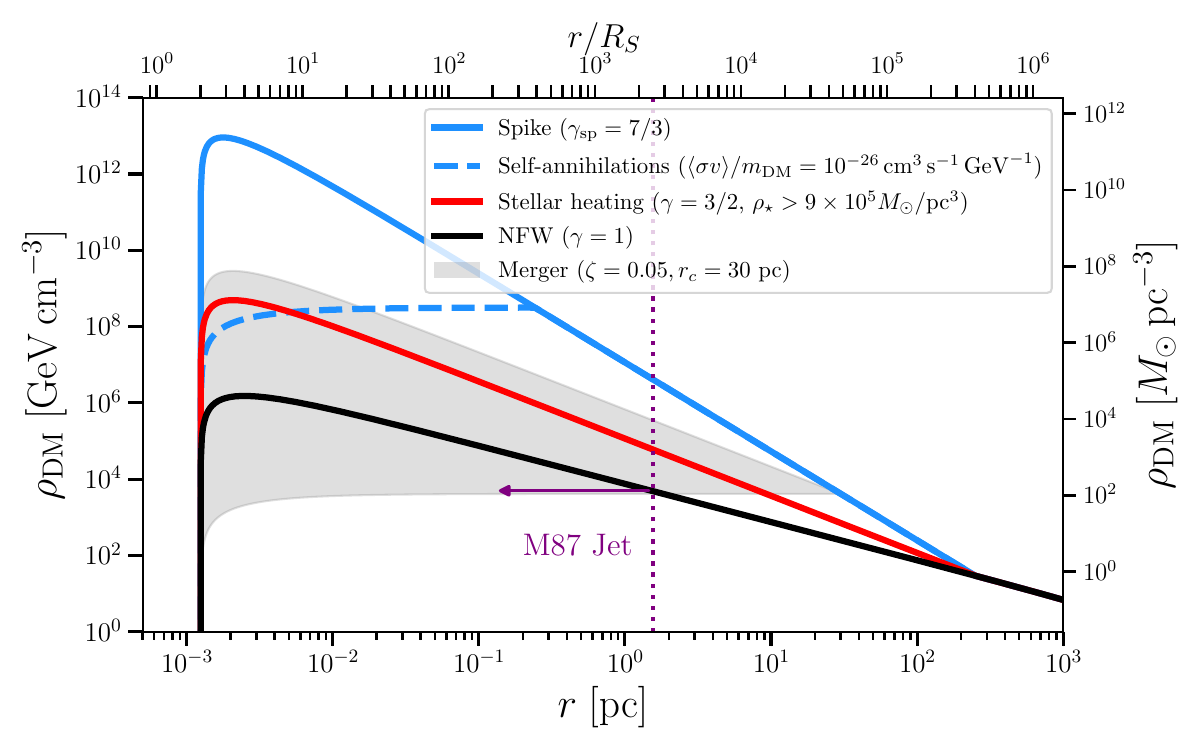}
    \caption{Possible dark matter density profiles in the inner parsecs of M87 under different astrophysical scenarios. The solid blue line corresponds to a canonical dark matter spike (see Eq. \ref{eq:spike}) formed under adiabatic growth of the SMBH and a pre-existing NFW profile with index $\gamma=1$ (also shown as a black solid line). The dashed blue line shows the depleted dark matter spike due to dark matter self-annihilation for a representative cross section value. The solid red line corresponds to a relaxed dark matter spike by stellar heating processes. We note that this effect is not expected to be efficient in M87 (see Appendix \ref{sec:appendix_DM} for details). The vertical dashed purple line indicates the edge of the M87 jet, taken at $r_{\rm jet}=2500R_S.$
    }
    \label{fig:density_profiles}
    
\end{figure}

Furthermore, M87's merger history plays a central role in shaping its present-day dark matter distribution. The formation of a supermassive black hole binary can rapidly evacuate stars and dark matter from the inner parsecs, removing mass comparable to that of the SMBH itself and further softening the spike index \cite{2006ApJ...648..976M, Merritt:2002vj}. M87 is expected to have undergone a merger between 1 and 10 Gyr ago \cite{2017ApJ...834...16M}, which could have softened the spike to an index in the range $\gamma_{\rm sp} = 0$–$1.5$ \cite{Merritt:2002vj, Merritt:2006mt} within the core radius $r_c$.

The size of the core can be estimated from the mass deficit induced by the merger \cite{Merritt:2002vj,Merritt:2006mt}. It has been discussed that the ejected mass is of order the SMBH mass per merger, i.e.
\begin{equation}
M_{\rm def} \;\simeq\; \sum_i \zeta(q_i)\,\big(M_{\bullet,1}^{(i)}+M_{\bullet,2}^{(i)}\big),
\end{equation}
where $M_\bullet$ is the mass of the SMBH, $q_i$ is the mass ratio of the $i$-th merger, and $\zeta(q)\sim 0.5$--$5$ parametrizes the efficiency of scouring. To translate $M_{\rm def}$ into a physical core radius $r_c$, we equate it to the pre-merger enclosed dark matter spike mass. The enclosed spike mass for the profile described in Eq. \ref{eq:spike} is
\begin{align}
M_{\rm sp}(<r) = 4\pi \rho_{R} R_{\rm sp}^{\gamma_{\rm sp}}
\int_{2R_S}^{r} g_{\gamma}(r')\, r'^{\,2-\gamma_{\rm sp}}\,dr'\equiv 4\pi \rho_{R} R_{\rm sp}^{\gamma_{\rm sp}}\, \mathcal{I}(r,\gamma_{\rm sp}),
\label{eq:enclosed_mass_general_fixed}
\end{align}
With the relativistic correction \(g_{\gamma}(r)\simeq (1-2R_S/r)\) one may write
\begin{align}
\mathcal{I}(r,\gamma_{\rm sp})=\int_{2R_S}^{r} \Big(1-\frac{2R_S}{r'}\Big)\, r'^{\,2-\gamma_{\rm sp}}\,dr'=\frac{r^{\,3-\gamma_{\rm sp}}-(2R_S)^{\,3-\gamma_{\rm sp}}}{3-\gamma_{\rm sp}}
-\frac{2R_S}{2-\gamma_{\rm sp}}\!\left[r^{\,2-\gamma_{\rm sp}}-(2R_S)^{\,2-\gamma_{\rm sp}}\right],
\label{eq:I_with_GR_fixed}
\end{align}
valid for \(\gamma_{\rm sp}\neq 2,3\).
If we further assume that the carved mass extends to a regime \(r\gg R_S\), the \(O(R_S/r)\) term is negligible and 
\begin{equation}
\mathcal{I}(r,\gamma_{\rm sp}) \;\simeq\; 
\frac{r^{\,3-\gamma_{\rm sp}}-(2R_S)^{\,3-\gamma_{\rm sp}}}{3-\gamma_{\rm sp}}.
\label{eq:I_noGR_fixed}
\end{equation}

We can then find \(r_c\) from the previously mentioned condition $M_{\mathrm{sp}}\left(<r_c\right)=M_{\mathrm{def}}$, leading to
\begin{align}
r_c 
&\simeq \Bigg[
(2R_S)^{\,3-\gamma_{\rm sp}} + 
\frac{(3-\gamma_{\rm sp})\,\zeta(q) M_{\bullet}}{4\pi \rho_{R} R_{\rm sp}^{\gamma_{\rm sp}}}
\Bigg]^{\!1/(3-\gamma_{\rm sp})},
\label{eq:rc_general_fixed}
\end{align}
where we have assumed \(g_\gamma\simeq\!1\), and a single merger with equal SMBH masses $M_{\bullet}$. 
This expression is fully determined by the spike parameters and the scouring efficiency. Numerically we find for M87 that $r_c \gtrsim R_{\rm sp}$ holds for a scouring efficiency $\zeta \gtrsim 0.2$, but for smaller values part of the original spike would survive; e.g for $\zeta=0.05$ the inner core is $r_c \simeq 30$ pc. 

In Fig. \ref{fig:density_profiles}, we show a variety of possible dark matter profiles in the vicinity of the SMBH of M87. The most pessimistic scenario corresponds to an efficient formation of an inner core from a recent SMBH merger with scouring efficiency $\zeta \gtrsim 0.05$. In this extreme case, we find that our limits on the dark matter-electron scattering cross section in Fig. \ref{fig:sigma_bound} would be relaxed by $\sim 6-7$ orders of magnitude, and the limits on the dark photon kinetic mixing from Fig. \ref{fig:t_sync} would be relaxed by a factor of $ \sim 3160$.

In the astrophysical scenarios considered to deplete the spike—whether through dark matter self-annihilation, stellar heating, or mergers—the bounds are relaxed by 2–3 orders of magnitude.
Nevertheless, the constraints continue to probe unexplored regions of the sub-GeV dark matter parameter space and surpass the reach of current direct-detection experiments.

\section{Uncertainties on the magnetic field of M87}
\label{sec:appendix_B}

\begin{figure}[H]
    \centering
    \includegraphics[width=0.7\linewidth]{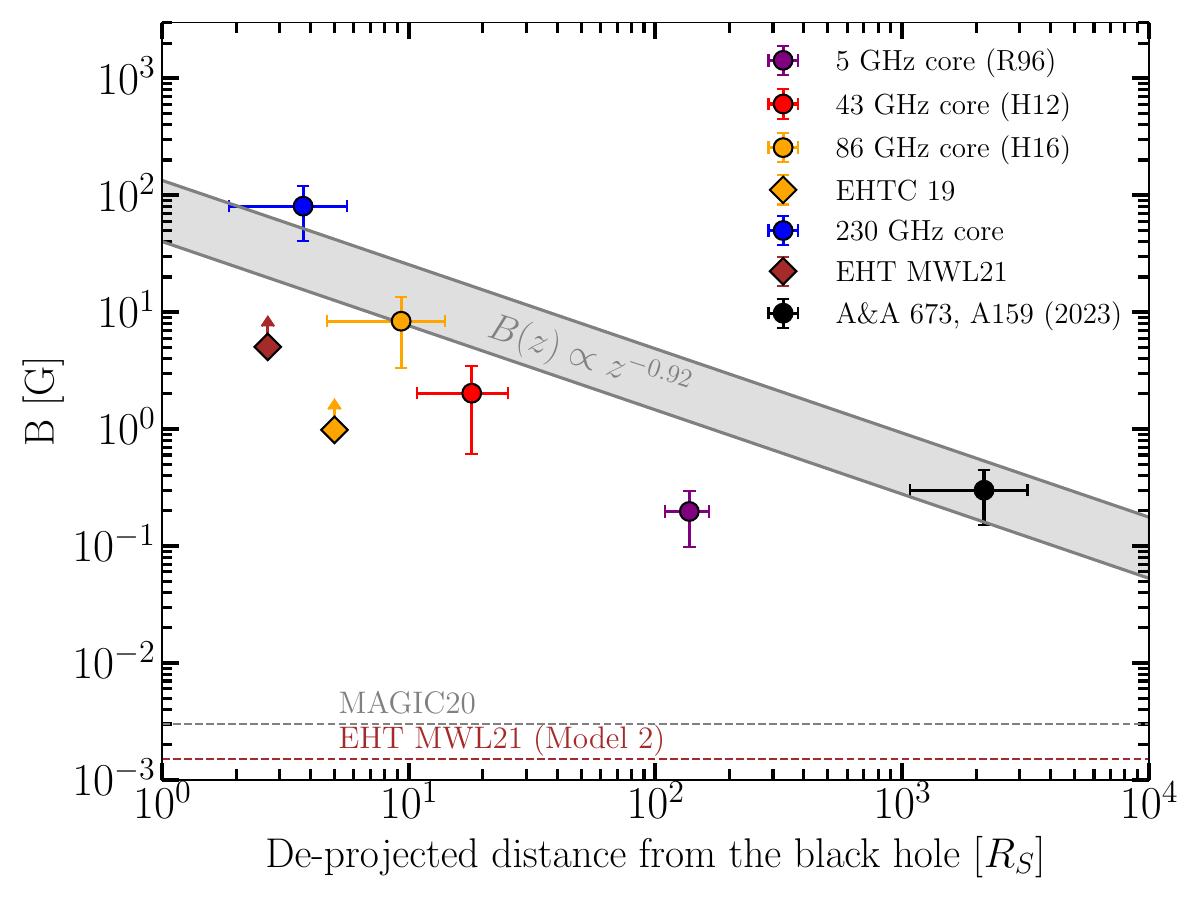}
    \caption{Magnetic field strength measurements and constraints of the M87 jet from VLBI cores at different frequencies (colored circle and diamond points)~\cite{Kino:2014dua, Kino:2015dha, Hada_2012, Hada_2016, Acciari_2010, Kim:2018hul, 2014Natur.510..126Z, EventHorizonTelescope:2019dse, Jiang_2021}, Spectral Energy Distribution (SED) fits at MAGIC (dashed gray line)~\cite{MAGIC:2020gbb}, and EHT measurements~\cite{EventHorizonTelescope:2021dvx}. The black point illustrates the estimated magnetic field strength from~\cite{Ro_2023}, and the gray region shows the extrapolated magnetic field from that study at different radii.}
    \label{fig:magnetic_field}
\end{figure}

The structure and strength of the magnetic field in M87’s jet remain subjects of significant uncertainty, with different observational methods probing distinct spatial scales and relying on various physical assumptions. In Fig. ~\ref{fig:magnetic_field} we show representative estimates and constraints spanning from the event-horizon of the central black hole up to the end of the jet of M87. Measurements based on VLBI core analyses at multiple frequencies (colored circles and diamonds) infer magnetic fields of order $B \sim 0.1$--$10~\mathrm{G}$ near the radio core through synchrotron self-absorption and core-shift techniques~\cite{Kino:2014dua, Kino:2015dha, Hada_2012, Hada_2016, Acciari_2010, Kim:2018hul, 2014Natur.510..126Z, EventHorizonTelescope:2019dse, Jiang_2021}. On larger scales, spectral energy distribution (SED) modeling of multi-wavelength emission provides complementary constraints, indicating magnetic field strengths of $B \lesssim 0.01$ G, as inferred from MAGIC observations (dashed gray line)~\cite{MAGIC:2020gbb}.
However, these estimates are subject to degeneracies between the particle density, Doppler boosting, and emission-zone size. The Event Horizon Telescope (EHT) polarimetric measurements further probe the magnetized plasma immediately surrounding the black hole, yielding estimates of $B \sim 1$--$10~\mathrm{G}$~\cite{EventHorizonTelescope:2021dvx}. Alternatively, Ref. \cite{Ro_2023} gives a measurement in the range $B \sim 0.1-1$G based on the spatial evolution of the synchrotron spectral index along the jet, further inferring a magnetic-field profile $B(z) \propto z^{-0.72}$. The shaded region in Fig.~\ref{fig:magnetic_field} illustrates the extrapolation of this result. The figure highlights the wide range of magnetic field strength estimates for M87; consequently, our bounds on the kinetic mixing and DM-electron cross section are bracketed by this uncertainty. We thus take an average of the VLBI core profile measurements, with a value of $\langle B_{\rm VLBI} \rangle =7$G, and the inferred value from the SED fits in MAGIC, of $B_{\rm MAGIC}= 0.003$G, and consider both cases in our analysis.

\section{Relationship between timescales and luminosity} \label{sec:optical_depth_argument}
It has been suggested that the nucleus of M87 is an order of magnitude less active than expected \cite{Reynolds_1996,prieto_2016}, and has led the field to developing new accretion models to explain the phenomena \cite{2000ApJ...539..809Q,1994ApJ...428L..13N,2014ARA&A..52..529Y,1994ApJ...428L..13N,Rees_1982}. In the main text we have investigated the possibility that DM-electron interactions reduce the expected luminosity from M87 to the observed level. If we attribute the difference between $L_{\text{Edd}}$ and $L_{\text{bol}}$ to DM cooling, we can relate the timescales as follows.  Given that the ratio of the observed $L_{\rm obs}$ and emitted $L_{\rm emit}$ fluxes, where $L_{\rm emit} = L_{\rm edd}$ and $L_{\rm obs} = L_{\rm bol}$ is
\begin{align}
    \frac{L_{\rm obs}}{L_{\rm emit}} &= \exp(-{n_{\rm DM} \sigma_{\rm DM-e} L})
\end{align}
where ${n_{\rm DM} \sigma L} = \frac{L}{l} =N$  where $N$ is the number of interactions, $l$ is the mean free path, and $L$ is the distance traveled along the jet. In the absence of DM-electron interactions, $L = \tau_{\rm sync} \ c$ in this context is the distance over which an electron radiates its synchrotron power. However, if DM-electron interactions are efficient, ${n_{\rm DM}} \sigma_{\rm DM-e} = (c \tau_{\mathrm{\chi}e})^{-1}$. Therefore
\begin{equation}
     \log\left (\frac{L_{\rm bol}}{L_{\rm edd}} \right ) = -\frac{\tau_{\rm sync}}{\tau_{\mathrm{\chi}e}}.
\end{equation}
From this ratio, we estimate that DM-electron interactions are efficient at depleting the Eddington luminosity to the observed level when $\tau_{\rm sync}/\tau_{\mathrm{\chi}e} \simeq 9$.

\end{document}